# Heterostructural one-unit-cell FeSe/SrTiO$_3$: from high-temperature superconductivity to topological states

Chaofei Liu[1], and Jian Wang[1,2,3,4†]

[1] International Center for Quantum Materials, School of Physics, Peking University, Beijing 100871, China
[2] Collaborative Innovation Center of Quantum Matter, Beijing 100871, China
[3] CAS Center for Excellence in Topological Quantum Computation, University of Chinese Academy of Sciences, Beijing 100190, China
[4] Beijing Academy of Quantum Information Sciences, Beijing 100193, China

[†] **E-mail:** jianwangphysics@pku.edu.cn

**Abstract** High-temperature superconductivity in one-unit-cell (1-UC) FeSe/SrTiO$_3$ heterostructure has become a research frontier in condensed-matter physics and material science. The superconducting transition temperature ($T_c$) of ultrathin FeSe film is significantly enhanced compared to its bulk counterpart and possibly approaches the liquid-nitrogen region according to *in situ* spectroscopic measurements. Particularly, the Fermi-surface topology of 1-UC FeSe consists of no hole pockets at Brillouin-zone center, which poses a great challenge to the well-accepted $s_\pm$-wave pairing nesting the sign-different electron–hole Fermi pockets in iron-based superconductors. In this review, we present the explorations of $T_c$ enhancement, electron pairing and topological phases in 1-UC FeSe/SrTiO$_3$. The potential coexistence of high-temperature superconductivity and topological electronic states promotes such two-dimensional heterostructure as the candidate of next-generation connate high-temperature topological superconductor and/or topological-quantum-computation platform.

## 1. Introduction

In 2008, LaO$_{1-x}$F$_x$FeAs was synthesized with a high superconducting (SC) transition temperature ($T_c$) of 26 K by Kamihara *et al*. [1], which pioneers the subsequent explosive 11-year studies of iron-based superconductors (IBSCs). The $T_c$ among iron-based bulk materials was shortly raised to 55 K [2], listing IBSCs in the family of high-temperature superconductors beyond cuprates. Different from the cuprates wherein the superconductivity is dominated by single Cu-$d_{x^2-y^2}$ orbital, the IBSCs involve five Fe-3$d$ orbitals ($d_{xy}$, $d_{xz}$, $d_{yz}$, $d_{x^2-y^2}$ and $d_{z^2}$) for the low-energy electronic properties. The multiorbital nature makes the pairing understanding of IBSCs complicated. Generally, the well-accepted pairing scenario in iron-based compounds is the interpockt $s$-wave pairing between the electron and sign-reversing hole Fermi pockets at Brillouin-zone (BZ) corners ($M$) and center ($\Gamma$), respectively, which is known as $s_\pm$ pairing [3,4]. However, the heavily electron-doped iron selenides (HEDISs) without $\Gamma$ hole pocket in the Fermi surface [5-8] make the initial $s_\pm$ scenario conceptually inapplicable, which stimulates intensive theoretical investigations on alternative pairing proposals [9-19].

Particularly, among the HEDISs, in 2012, Wang *et al*. discovered an impressively large SC-like gap (20 meV) in the interface-engineered one-unit-cell-thick (1-UC) FeSe film grown on SrTiO$_3$(001) substrate [20], which attracted great attention in the community of condensed-matter physics [21,22]. The direct evidence for superconductivity therein above 40 K ($T_c^{\text{onset}} \approx 54.5$ K) was subsequently given by Zhang and Sun *et al*. in 2014 via *ex situ* electrical-transport experiment [23]. Thus, the $T_c$ of FeSe at 1-UC limit is dramatically higher than that of bulk FeSe (8–9 K) and challenges the expectation of decreased $T_c$ with reduced lateral dimension. Angle-resolved photoemission spectroscopy (ARPES) studies reveal that the gap-disappearing temperature ($T_g$; denoted as $T_c$ hereafter for simplicity) is typically 65 K [8,24-27], which indicates the 1-UC FeSe may show the highest $T_c$ in IBSCs. The *in situ* four-point probe electrical-transport [28] and *ex situ* magnetic-susceptibility measurements [29] even suggest the $T_c$ might be higher than liquid-nitrogen temperature. Significant efforts have been devoted to disentangling various interface effects for explaining the unexpected colossal $T_c$ enhancement [24,30-37]. The charge transfer across the FeSe/SrTiO$_3$ interface has been believed to be especially vital for the detected high $T_c$. Furthermore, among the IBSCs, the binary compound FeSe has the simplest crystal structure and is the building block of iron chalocogenides. The molecular-beam-epitaxy technique utilized for ultrathin-FeSe growth further guarantees the ultrahigh sample quality at atomic level. Therefore, 1-UC FeSe/SrTiO$_3$ provides an ideal platform for both theoretical and experimental studies upon the pairing mechanism and SC properties for high-$T_c$ superconductors.

In cuprates and heavy-fermion superconductors, extensive explorations reveal that the unconventional superconductivity emerges in proximity to the antiferromagnetically ordered parent state in the temperature–carrier doping phase diagram. Provided the antiferromagnetically [38-42] insulating [43] or semimetallic parent phase [44] of 1-UC FeSe, it would be interesting to see whether the high-temperature superconductivity among different compounds showing varied details shares the same SC mechanism. Besides the interface-engineered high-temperature superconductivity, other exotic physics may occur in 1-UC FeSe under different conditions, including electron correlations [45,46], orbital-selective Mott phase [46], insulator–superconductor crossover [43], bosonic-type excitations [47,48] and nematic fluctuations [49,50]. Especially, the topological phases addressed in 1-UC FeSe [41,44,51-53] may render connate high-temperature topological superconductivity and Majorana bound states (MBSs) at two-dimensional (2D) limit.

In this review, we briefly summarize the current understanding of high-temperature superconductivity and topological electronics in 1-UC FeSe/SrTiO$_3$. In Part 2, we discuss the influence of several interface effects on $T_c$. In Parts 3 and 4, we introduce the possible pairing mechanism of 1-UC FeSe and make a comparison with other HEDISs. In Part 5, we summarize the progress in exploring the topological phases in 1-UC FeSe. Finally, in Part 6, we conclude with the perspectives.

## 2. Understanding of $T_c$ enhancement

The dramatic $T_c$ enhancement in 1-UC FeSe/SrTiO$_3$ with respect to the bulk material is attributed to the interface effects. From the fundamental perspective, the interface engineering in heterostructure can hybridize the electron wavefunctions of different materials and likely results in exotic quantum phenomena. One of the representative examples is the heterointerface of two insulating perovskites, LaAlO$_3$(001) and SrTiO$_3$(001), which shows ultrahigh carrier mobility (> 10,000 cm$^2$V$^{-1}$s$^{-1}$) and unexpected interface superconductivity ($T_c \approx 200$ mK) [54]. For heterostructural FeSe/SrTiO$_3$, the interface effects, including electron doping, tensile strain, and influence of Se−Fe−Se bond angle, dielectric constant and double-TiO$_2$ termination, have been found to tune the $T_c$ in different degrees. We now address this issue in detail as follows.

### 2.1 Electron doping

In the weakly coupled Bardeen–Cooper–Schrieffer (BCS) formulism [55],

$$T_c = 1.14 w_D \exp\left[-\frac{1}{N(0)V}\right], \tag{1}$$

where $w_D$ is the Debye temperature, $N(0)$ is the density of states at Fermi energy ($E_F$) and $V$ is the pairing potential arising from electron–phonon coupling (EPC). According to equation (1), $T_c$ is positively correlated to $N(0)$, meaning charge density is beneficial to the strengthening of superconductivity. In experiments, electron doping in FeSe film can be realized by thermal annealing, surface dosing of alkali-metal atoms and voltage gating. The maximum doping level typically reaches 0.12 $e^-$/Fe and can solely drive $T_c$ up to a magnitude of ~50 K.

*Oxygen vacancies and annealing.*—The oxygen vacancies on the top layer of SrTiO$_3$ [25] are the intrinsic donors to dope electrons into 1-UC FeSe. The vacancies are introduced during the substrate-treatment procedure by Se-flux etching at high temperatures (typically 950 °C) before the film growth. To compensate for the work-function difference across the FeSe/SrTiO$_3$ interface, the negative charges from oxygen vacancies tend to be transferred to FeSe. Physically, the transferred carriers will fill the hole pocket at Γ point of 1-UC FeSe and provide strong Coulomb binding between FeSe layer and SrTiO$_3$ substrate [56]. As the Fermi levels of both systems are aligned in a new equilibrium state, the charge transferring terminates and induces a band bending at the heterostructure interface, which has been captured by both electron-energy-loss spectroscopy [57] and X-ray photoemission spectroscopy [58].

Successive thermal annealing in ultrahigh vacuum facilitates the charge-transferring process by oxygen loss in SrTiO$_3$ substrate [24]. In fact, the as-grown ultrathin FeSe sample is insulating and superconductivity emerges by post-annealing [59]. During the insulator–superconductor crossover as annealing, the hole band at Γ point gradually sinks below $E_F$, and both the Fermi surface and the electronic dispersions experience a systematic evolution. Meanwhile, the electron-doping level, which can be estimated from the size of Fermi pockets around $M$ points, increases and saturates at 0.12 $e^-$/Fe at the optimized annealing condition [figure 1(a1)]. Correspondingly, the maximum $T_c$ reaches 65 K [figure 1(a2)] as indicated by the gap-vanishing temperature [24]. Note that the 0.08-$e^-$/Fe–doped, (Li$_{1-x}$Fe$_x$)OH-intercalated bulk FeSe [(Li$_{0.84}$Fe$_{0.16}$)OHFe$_{0.98}$Se] without interface effects also shows enhanced $T_c$ of 41 K [60] compared with the pure FeSe crystal. Thus, the electron doping is probably the major origin of the dramatic $T_c$ enhancement in 1-UC FeSe. Since Se atoms will desorb with annealing and also introduce extra doping electrons, the identification of the leading carrier contribution still requires further investigations.

*K-atom dosing.*—For multilayer FeSe film on SrTiO$_3$ substrate, while the 1$^{st}$ UC remains SC, the layers ≥ 2$^{nd}$ UC show no signatures of superconductivity. The SC states selectively limited within the 1$^{st}$ UC suggest the essential role of the interface effects in tuning the high-temperature superconductivity in 1-UC FeSe/SrTiO$_3$. A direct method to check whether the electron transferring behaves as the dominating interface effect to enhance $T_c$ is dosing alkali-metal atoms on the top surface of multilayer FeSe film. Experimentally, consecutive *in situ* dosing of K atoms onto 3-UC FeSe/SrTiO$_3$ can indeed convert the non-SC film into a superconductor [30]. Thus, the disappearance of SC states in bare multilayer FeSe film is largely attributed to the insufficient electron doping from the substrate. Different from the 1-UC sample, the undoped 3-UC film shows the Fermi-surface topology consisting of a hole pocket (*α*) at Γ point and a small electron pocket (*ε*) near $M$ point. As increasing the K-atom dosage, a new electron pocket (*γ*) emerges at $M$ point, accompanied by the downward shift of the hole band at Γ point and accordingly, the disappearance of *α* pocket. Upon further increasing the K dosage, the *α* band continues shifting downwards and the volume of *γ* pocket gradually increases. Simultaneously, the SC gap firstly increases and later decreases towards the heavily electron-doped region (0.15 $e^-$/Fe), showing a dome-like SC phase [figure 1(b)]. The optimal doping is achieved at 0.11 $e^-$/Fe, corresponding to $T_c$ of 48 K [30], which is slightly lower than the interface-engineered $T_c$ of 65 K in 1-UC FeSe/SrTiO$_3$. The $T_c$ difference (65 K vs. 48 K) indicates the interface effects besides electron transfer can also contribute to the observed $T_c$ enhancement in 1-UC FeSe/SrTiO$_3$.

*Gate-voltage tuning.*—The ionic-liquid gating in an electric-double-layer transistor is an alternative technique for efficiently tuning the carrier density. In FeSe thin films and flakes without epitaxial interfaces, the $T_c$ can be profoundly increased by liquid gating [31,61,62]. For example, before applying the gate voltage ($V_g$), the FeSe thin flake (typically 10-nm-thick) [31] shows $T_c$ of 5.2 K, slightly lower than that of the bulk material (8 K). As increasing the $V_g$, the $T_c$ continuously increases and can reach as high as 45.3 K at the maximum $V_g$ (6 V) available in experiments. The $T_c$ at 6-V $V_g$ is sample-dependent and the highest value achieved is 48 K. In the $V_g$-tuned high-$T_c$ phase, the Hall coefficient is negative, corresponding to the dominant electron-type carriers. The detailed $V_g$ evolution of $T_c$ and the carrier density determined via Hall effect, $n_H$, are summarized in figure 1(c). It is seen that, upon applying $V_g$, as $T_c$ is slowly enhanced, $n_H$ initially decreases because of electron doping. At the critical $V_g$ of 4.25 V, where the high-$T_c$ state emerges, $n_H$ suddenly drops with a sign reversal. As further increasing $V_g$, whereas the $T_c$ rapidly ramps up, $|n_H|$ decreases synchronously. The high tunability of $T_c$ towards 48 K by gate voltage in FeSe flakes again implies the electron doping as the main origin of high-temperature superconductivity in 1-UC FeSe/SrTiO$_3$.

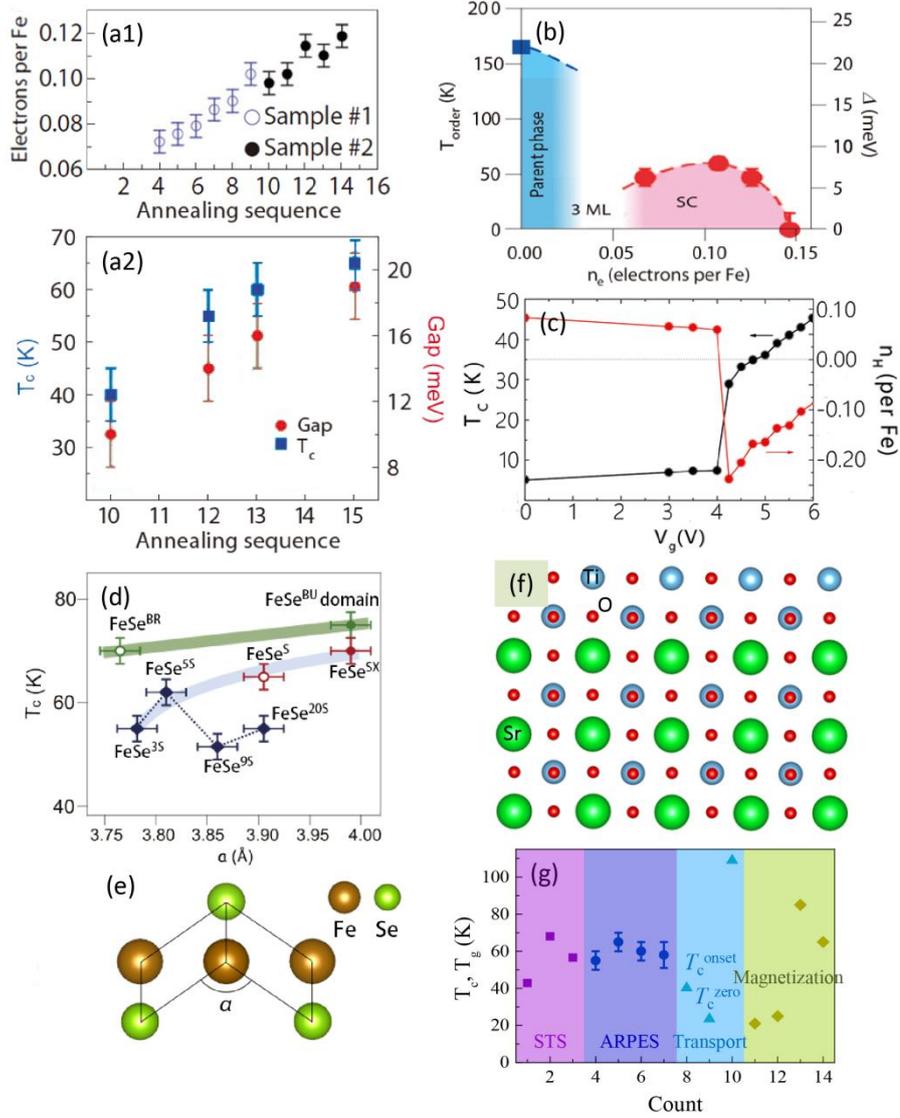

**Figure 1.** Tuning $T_c$ of 1-UC FeSe film by electron doping, tensile strain, Se−Fe−Se bond angle and double-TiO$_x$ termination. (a1,a2) Electron doping and $T_c$ in different annealing steps of 1-UC FeSe. (b) SC gap as a function of electron doping tuned by K-atom dosing on 3-UC FeSe. (c) $V_g$ dependence of $T_c$ and $n_H$ for FeSe thin flake. (d) $T_c$ plotted vs. in-plane lattice constant $a$ of 1-UC FeSe. (e,f) Schematics of Se−Fe−Se bond angle and double-TiO$_x$ termination of SrTiO$_3$. (g) $T_c$ ($T_g$) measured by different experimental techniques [8,20,23-25,28,29,47,48,59,63]. Reproduced from [24,30,31,34].

**2.2 Tensile strain**

1-UC FeSe (lattice constant, $a$ = 3.82 Å [23]) on SrTiO$_3$ is lattice-expanded by 1.5% compared with bulk FeSe (3.765 Å [64]). In the generalized gradient approximation incorporating local Coulomb repulsions (GGA+$U$), the tensile strain exerted on 1-UC FeSe by lattice expansion can stabilize the collinear antiferromagnetic (AFM) state in Fe layer by increasing the superexchange interactions between the next-nearest Fe atoms, which are intimately correlated with the high $T_c$ in 1-UC FeSe [65]. In experiments, interface engineering by using different substrates provides a practically accessible method to probe the tensile-strain effect on superconductivity. Accordingly, highly strained 1-UC FeSe (FeSe$^X$) was synthesized on Nb:SrTiO$_3$/KTaO$_3$ heterostructure, which was designed to increase the tensile strain in FeSe film while preserving the FeSe/Nb:SrTiO$_3$ interface [32]. The lattice constant of 1-UC FeSe is prominently expanded to 3.99 ± 0.02 Å by heterostructural Nb:SrTiO$_3$/KTaO$_3$ substrate with the lattice constant of 3.989 Å. Near Γ point, the band dispersions include a parabolic hole band, $α$, and a relatively flat band, $ω$, at higher binding energy both below $E_F$. Compared with moderately strained 1-UC FeSe (FeSe$^M$) on "normal" Nb:SrTiO$_3$ substrate ($a$ = 3.905 Å), FeSe$^X$ shows similar Fermi-surface topology without the hole pockets at Γ point and nodeless, anisotropic SC-gap distribution with C4 symmetry in reciprocal space. Nevertheless, the ellipticity of the two electron Fermi pockets at $M$ points for FeSe$^X$ is increased by the extreme tensile strain without observable hybridization. Particularly, the $T_c$ defined by gap-disappearing temperature in FeSe$^X$ persists only to 70 K. The electron doping estimated from the Fermi-surface volume is 0.12 $e^-$/Fe, equal to that of FeSe$^M$ with $T_c$ typically of 65 K by the same definition. With the same charge concentration as FeSe$^M$, the only 5-K $T_c$ enhancement in FeSe$^X$ by 4.5% lattice expansion [3.99 Å (FeSe$^X$) vs. 3.82 Å (FeSe$^M$)] indicates a negligible effect of the tensile strain on $T_c$ within the resolution of ARPES experiments. In contrast, the increasing of $T_c$ from 8 K in bulk FeSe to 65 K in FeSe$^M$ with 1.5% lattice expansion is comparably rather large, suggesting other interfacial factors beyond the tensile strain, likely electron doping, dominate the interfacial-superconductivity enhancement.

Investigations of 1-UC FeSe on the substrates showing different lattice constants have tremendously extended the tensile-strain range. The interface lattice control includes heterostructure of Nb:BaTiO$_3$/KTaO$_3$ with unrotated (FeSe$^{BU}$, $a$ = 3.99 Å) and rotated above-grown FeSe lattices (FeSe$^{BR}$, $a$ = 3.78 Å) (both doped by 0.12 $e^-$/Fe) [33], as well as thickness-varied $n$-UC Nb:SrTiO$_3$/LaAlO$_3$ (FeSe$^{nS}$, $a$ = 3.79–3.9 Å) [34]. Compared with the calculated bands of free-standing 1-UC FeSe by density functional theory (DFT), the band structures of differently strained 1-UC FeSe revealed by ARPES are all renormalized by the interfacial oxides. Correspondingly, the effective electron mass of $α$ and $γ$ bands at Γ and $M$ points, respectively, increases as the lattice of 1-UC FeSe expands. Both the band renormalization and the electron-mass variation suggest the tensile strain in these interfacial systems with different substrates can tune the electron correlations in FeSe films. More intriguingly, as summarized in figure 1(d), the $T_c$ is weakly and positively correlated to the tensile strain via tuning the lattice constant in 1-UC FeSe in statistics. The statistically positive trend in $T_c$–tensile strain diagram starts nearly from the bulk-FeSe lattice constant of 3.765 Å. Near the 3.765-Å lattice region, the FeSe$^{BR}$ and FeSe$^{3S}$ are negligibly enlarged compared with bulk FeSe lattice but both show enhanced $T_c \geq 55$ K (70 and 55 K, respectively). The robustness of high-temperature superconductivity approaching the regime for the lattice constant of bulk FeSe further manifests that the tensile strain is only a positive factor for the $T_c$ enhancement in strained 1-UC FeSe, but not the major factor. Compared with FeSe$^M$ showing moderate tensile strain, FeSe$^{3S}$ is exerted with relatively negligible tensile strain. Their $T_c$ contrast (65 K vs. 55 K) implies the tensile strain facilitates $T_c$ by a magnitude of ~10 K.

## 2.3 Se−Fe−Se bond angle

Empirically, in experiments about the bulk materials of iron chalocogenides, larger X−Fe−X (X = S, Se, Te) bond angle [figure 1(e)], or equivalently, lower chalcogen height with respect to Fe layer, corresponds higher $T_c$ [66]. In previous studies, both the bond angle and the chalcogen height were indirectly calculated from the X-ray diffraction data of iron-chalocogenide compounds. In 1-UC FeSe, these parameters have been directly determined by the high-resolution scanning transmission electron microscopy (STEM) [35], yielding accurate estimation of their correlations with superconductivity. Specifically, as 1-UC FeSe evolves from non-SC to SC states by successive annealing, the Se–Fe–Se angle monotonously increases from 110.0 ° ± 0.9 ° to 111.4 ° ± 0.9 °. Meanwhile, the Se height decreases from 1.33 ± 0.02 Å to 1.31 ± 0.01 Å. Both phenomena reveal a close correlation between the interface structure and superconductivity for 1-UC FeSe/SrTiO$_3$. Because the effect of the Se–Fe–Se angle on superconductivity is entangled with other factors, like charge transfer and tensile strain, the contribution of bond angle to $T_c$ cannot be quantified.

**2.4 Dielectric constant**

While 1-UC FeSe/SrTiO$_3$ shows a high $T_c$ typically of 60–65 K by STM and ARPES, 1-UC FeSe/SiC is non-SC down to 2.2 K [67]. Note that SrTiO$_3$ possesses a much higher dielectric constant (~1 × 10$^4$) than that of SiC (~10) at low temperatures (~1 K). The contrast of $T_c$ in 1-UC FeSe on SrTiO$_3$ and SiC possibly signifies the necessary role of high dielectric constant for the emergence of interface-engineered high-temperature superconductivity. Rutile TiO$_2$(100) substrate is another choice for epitaxial ultrathin-film synthesis, where the atomic flatness over macroscopic scale has been successfully realized [68]. Control experiments of the dielectric constant by ARPES have been performed on 1-UC FeSe film grown on rutile TiO$_2$(100) [36]. The spectroscopic results demonstrate the detected electronic structures, SC-gap size and $T_c$ all highly resemble those of 1-UC FeSe/SrTiO$_3$. Given that the low-temperature dielectric constant of rutile TiO$_2$ is only ≤ 260, much smaller than that of SrTiO$_3$, the "anomalous" similarities between 1-UC FeSe on SrTiO$_3$ and TiO$_2$ likely exclude the importance of the dielectric constant to the $T_c$ enhancement. More systematic investigations are required to clarify the role of the dielectric constant in affecting the interface superconductivity.

**2.5 Double-TiO$_x$ termination**

Single-TiO$_2$–terminated SrTiO$_3$ was commonly assumed in previous band calculations for 1-UC FeSe/SrTiO$_3$. Nevertheless, STEM suggests the SrTiO$_3$ with above-grown 1-UC FeSe is terminated with the double-TiO$_x$ layer [figure 1(f)] [35]. Specifically, in the atomically resolved high-angle annular-dark-field image of FeSe/SrTiO$_3$ interface, an extra layer is detected between the bottom Se layer of 1-UC FeSe and the top TiO$_2$ layer of SrTiO$_3$. The atom columns of this extra layer exhibit the intensities comparable with those of Ti, directly indicating the feature of TiO$_x$ layer. Quantitatively, using synchrotron X-ray diffraction, the crystal truncation rod along (2, 0, $l$) of SrTiO$_3$ is well fitted by the model of double-TiO$_x$ surface termination [37], also demonstrating the double-TiO$_x$ termination of SrTiO$_3$. Due to the presence of the double-TiO$_x$ later, the inter-TiO$_x$-layer distance is lengthened by 25%, and both the 1$^{st}$ TiO$_x$ and the 1$^{st}$ SrO layers [2$^{nd}$ and 3$^{rd}$ layers from top to bottom in figure 1(f), respectively] are shrinked downwards accordingly.

The double-TiO$_x$ interface facilitates the electron transfer from SrTiO$_3$ to FeSe to strengthen superconductivity by modifying the electronic structures [37]. DFT calculations show that the double-TiO$_x$–terminated SrTiO$_3$ with 50% oxygen vacancies more strongly bonds to FeSe compared with the single-TiO$_x$ termination case. Consequently, the double-TiO$_x$–terminated interface promotes the charge transfer more effectively than the single-TiO$_x$ termination, which submerges the hole pocket at Γ point below $E_F$ and reproduces the band structures as observed in ARPES.

In practical situations, the ingredients of the interface effects are mutually entangled and one of them is hard to be fully decoupled from the others for more precise analysis. Despite of these logical difficulties, based on all above discussions, the dramatic $T_c$ enhancement in 1-UC FeSe [summarized in figure 1(g) obtained by different techniques] can be qualitatively explained by the collaborating effect of charge doping/transfer (dominated) and tensile strain (secondary). The remaining components of the interface effects are either incorporated into the influence of charge doping and tensile strain, or negligible in $T_c$ enhancement.

## 3. Pairing insights

BCS theory predicts that superconductivity emerges via the condensation of Cooper pairs to the ground state. The identical-particle effect dictates that the wavefunction of Cooper-paired electrons,

$$\psi(s_1, s_2; \boldsymbol{r}_1, \boldsymbol{r}_2) = \chi(s_1, s_2)\phi(\boldsymbol{r}_1, \boldsymbol{r}_2), \quad (2)$$

be antisymmetric, where $\chi$ and $\phi$ are spin and orbital wavefunctions, respectively. Accordingly, the spin-singlet state,

$$\chi_s(s_1, s_2) = \frac{1}{\sqrt{2}}\left[\chi_{\frac{1}{2}}(s_1)\chi_{-\frac{1}{2}}(s_2) - \chi_{\frac{1}{2}}(s_2)\chi_{-\frac{1}{2}}(s_1)\right], \quad (3)$$

relates to symmetric orbital wavefunction (even-parity), $\frac{1}{\sqrt{2}}[\phi_1(\boldsymbol{r}_1)\phi_2(\boldsymbol{r}_2) + \phi_1(\boldsymbol{r}_2)\phi_2(\boldsymbol{r}_1)]$, while the spin-triplet state,

$$\chi_t^1(s_1, s_2) = \chi_{-\frac{1}{2}}(s_1)\chi_{-\frac{1}{2}}(s_2),$$

$$\chi_t^2(s_1, s_2) = \chi_{\frac{1}{2}}(s_1)\chi_{\frac{1}{2}}(s_2), \quad (4)$$

$$\chi_t^3(s_1, s_2) = \frac{1}{\sqrt{2}}\left[\chi_{\frac{1}{2}}(s_1)\chi_{-\frac{1}{2}}(s_2) + \chi_{\frac{1}{2}}(s_2)\chi_{-\frac{1}{2}}(s_1)\right],$$

relates to antisymmetric orbital wavefunction (odd-parity), $\frac{1}{\sqrt{2}}[\phi_1(\boldsymbol{r}_1)\phi_2(\boldsymbol{r}_2) - \phi_1(\boldsymbol{r}_2)\phi_2(\boldsymbol{r}_1)]$. The quantum numbers of the orbital angular momenta of Cooper pairs, $L = 0$, $\hbar$ and $2\hbar$ ($\hbar$ being reduced Planck constant), are termed as *s*-, *p*- and *d*-wave, which are known as the Cooper-pairing symmetry of superconductors. Given the orbital-wavefunction symmetry, the *s*- and *d*-wave states with even angular momenta correspond to spin-singlet pairings, while the *p*-wave with odd angular momentum corresponds to spin-triplet pairing. The SC order parameter (OP) is described by $\Delta = \Delta_0 e^{i\theta}$, where $\Delta_0$ and $\theta$ represent magnitude and phase, respectively. In high-temperature superconductors, the pairing mechanism in SC ground states, e.g., pairing symmetry and the structure of SC OP, is still under intensive debate.

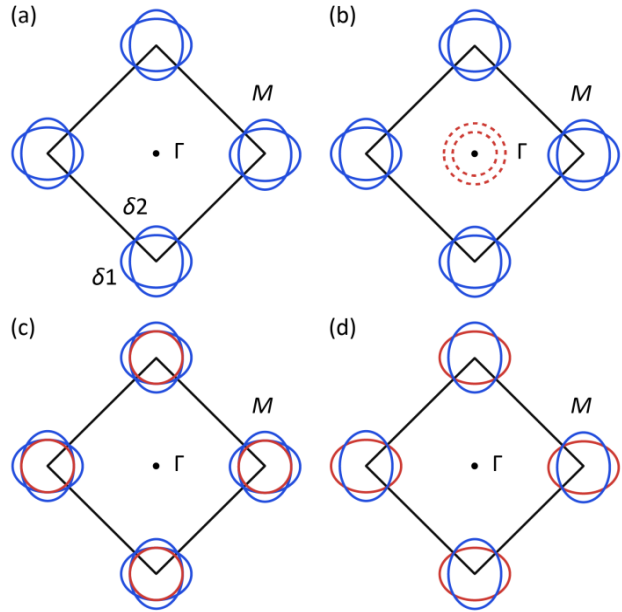

**Figure 2.** Fermi-surface topologies of 1-UC FeSe in folded BZs for (a) anisotropic $s_{++}$-, (b) incipient $s_{\pm}$-, (c) extended $s_{\pm}$- and (d) nodeless *d*-wave pairings, respectively. SC OPs of opposite signs on Fermi sheets are shown in different colors. The two ellipse Fermi pockets at each *M* point are in reality hybridized at intersections to avoid crossings.

Glide–mirror-symmetry breaking in 1-UC FeSe due to the interface engineering by $SrTiO_3$ doubles the 1-Fe unit cell and folds the BZ. Hence, the horizontal ellipse electron pocket ($\delta_1$) overlaps onto the vertical counterpart ($\delta_2$) at M points in the folded BZ [e.g., figure 2(a)]. The presence of spin–orbital coupling (SOC) splits the band degeneracy at hybridized $\delta_1/\delta_2$ crossings, resulting in the nodeless SC-gap function [8]. Different from iron pnictides, 1-UC FeSe/$SrTiO_3$ consists of only electron pockets at BZ corners [8], whereas the hole pockets at BZ center sink below $E_F$ due to the heavy electron doping. In consequence, the prevailing $s_\pm$-wave pairing based on electron–hole Fermi-pocket nesting [3,4] in iron pnictides is challenged. The absence of Γ hole pockets, together with the nodeless SC gap, puts considerable constraints on the pairing scenarios of 1-UC FeSe. Regarding the pairing symmetry and the sign structure, the hole-pocket–absent Fermi-surface topology of 1-UC FeSe yields sign-preserving plain $s_{++}$- [9], anisotropic $s_{++}$- [10,11], and sign-reversing "incipient" $s_\pm$- [13-15], extended $s_\pm$- [16,17] and nodeless *d*-wave pairings [19] (figure 2). More detailed calculations additionally proposed orbital-fluctuation–mediated $s_{++}$- [11,12] and parity-mixed *s*-wave [18] pairing possibilities.

### 3.1 Sign-preserving pairings

*Plain $s_{++}$-wave.*—Moderate $\delta_1/\delta_2$ hybridization makes the electron pockets at M points tend to be circular. The resultant pairing that preserves the sign of gap function is plain $s_{++}$-wave. Under an effective Hamiltonian approach, the magnetic exchange couplings, $J_{ij}$, are truncated to only the nearest- ($J_1$), or the next-nearest-neighbor AFM interactions ($J_2$) [9]. In $J_1$–$J_2$ phase diagram, the $s_{++}$-wave pairing dominates with increased stability by occupying a leading area.

Quasiparticle interference (QPI) and impurity scatterings are both phase-sensitive methods for probing the pairing symmetry and the sign structure of superconductors. For QPI, the scattered Bogoliubov quasiparticles off scattering centers constructively interfere, inducing spatially periodic modulations of density of states. By fast Fourier transformation (FFT), the QPI patterns sketch the contours of scattering vectors in *q* space [69]. Here, $q = k_f − k_i$, where $k_i$ and $k_f$ are the momenta of initial and final states, respectively, in a scattering process. In 1-UC FeSe, the systematic QPI analysis shows similar energy dependence of intra- and interpocket scattering intensities near the SC-gap edge before and after applying magnetic field [figures 3(a)–(c)] [70]. These uniform energy dispersions directly suggest the equivalence of all scattering channels and presumably, the equivalence of the Fermi pockets at different M points. Within unfolded-BZ framework, the nodeless *d*-wave is excluded, where the electron pockets at adjacent M points are inequivalent with sign reversal.

The response of superconductivity to local impurities sensitively depends on the pairing symmetry and the impurity characteristics (table 1). Different from magnetic impurities that are pairbreakers for both sign-preserving and sign-reversing pairings owing to time–reversal-symmetry breaking, nonmagnetic impurities selectively destroy $s_\pm$-, *p*- and *d*-wave pairing states by in-gap quasiparticle bound states [71]. The $s_{++}$-wave is preserved by Anderson's theorem [72], which is barely suppressed by nonmagnetic scatterings. Some scanning-tunneling-spectroscopy (STS) experiments reported the absence of sharp in-gap bound states on nonmagnetic impurities on/in 1-UC FeSe [figures 3(d1)–(e2)], including Zn, Ag, K adatoms [70] and Fe vacancy [73]. In combination with the robust local superconductivity of 1-UC FeSe against severe K-adatom disorders [74], plain $s_{++}$-wave is suggested by these investigations likely as the pairing symmetry of 1-UC FeSe.

However, identifying the nonmagnetic-impurity effects is generally nontrivial in experiments. Especially for weak and extended nonmagnetic scattering potentials, the intrapocket scatterings are mainly induced, which do not contribute to the formation of in-gap bound states in multiband sign-reversing superconductivity. Actually, the $s_{++}$-wave pairing in 1-UC FeSe concluded from nonmagnetic scatterings has received theoretical suspicion [75]. Within the standard *T*-matrix approximation, a phenomenological theory of impurity scatterings shows that the impurity bound states saturating at the SC-gap edge will be effectively unobservable for sign-reversing incipient $s_{\pm}$-wave [75], which behave experimentally indistinguishable from the $s_{++}$-pairing case without bound states. Scrutinization of the tunneling spectra taken upon the reported nonmagnetic impurities (e.g., Ag, K adatoms and Fe vacancy) indeed reveals SC-gap suppression and electron- or hole-like spectral-weight enhancement compared to the "normal" SC spectra [figures 3(e1)–(e2)]. Clarifying whether the weak spectral-reconstruction features signal the saturated bound states towards SC-gap edge is crucial in decisively determining the pairing scenario of 1-UC FeSe.

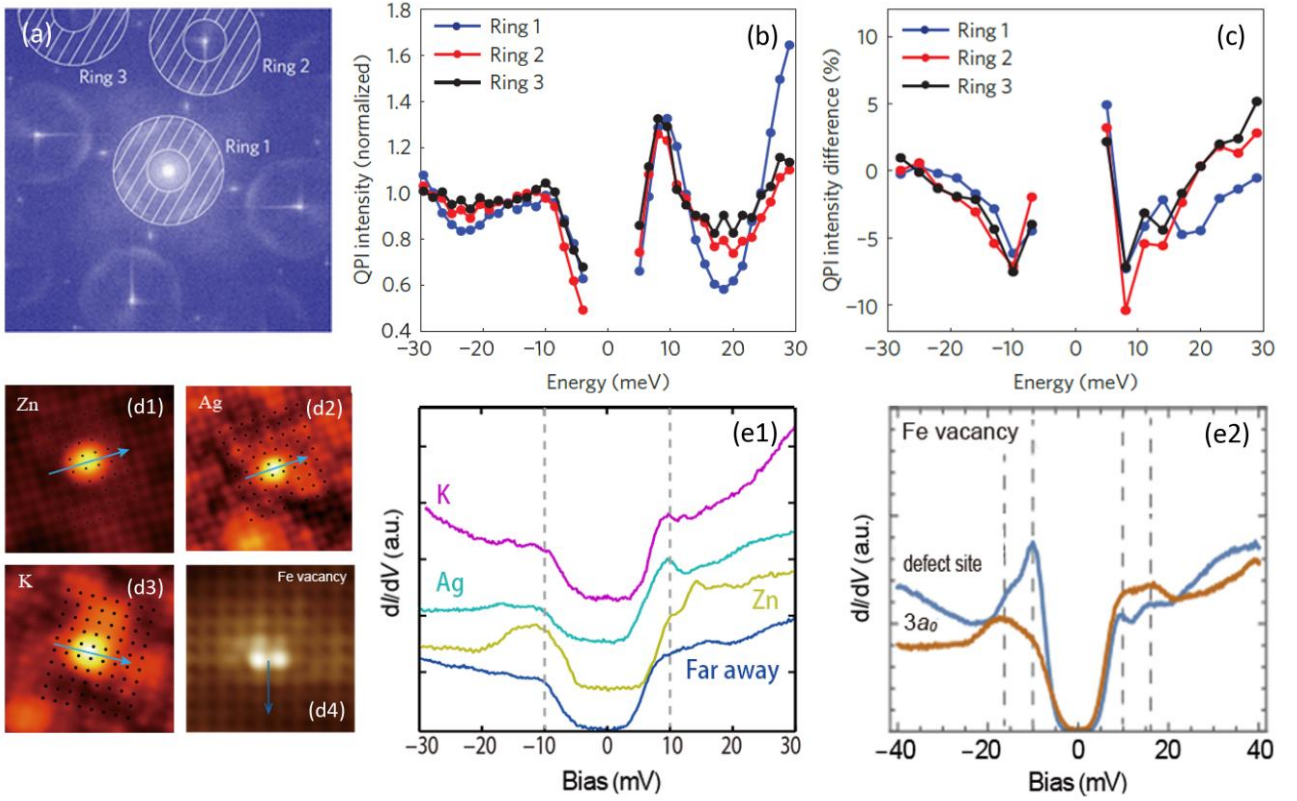

**Figure 3.** QPI and impurity-scattering analyses of 1-UC FeSe. (a) QPI pattern at 16.5 meV. (b) Intensities of intra- [$q_1$ (ring 1)] and interpocket scatterings [$q_2$, $q_3$ (rings 2, 3)] as functions of energy. (c) Difference between intra-/interpocket scattering intensities at $B = 11$ T and $B = 0$ T. (d1–d4) Zn, Ag, K adatoms and Fe vacancy, and (e1–e2) corresponding tunneling spectra. Reproduced from [70,73].

**Table 1.** Impurity-scattering effects for different pairings. "√" and "×" denote that the impurity scattering induces and does not induce local pairing breaking, respectively.

| Pairing symmetry | Sign-preserving pairing | Sign-reversing pairing | | |
|---|---|---|---|---|
| | $s$ | $s_{\pm}$ | $p$ | $d$ |
| Nonmagnetic impurity | × | √ | √ | √ |
| Magnetic impurity | √ | √ | √ | √ |

According to QPI selection rules (table 2), magnetic impurities, which are odd under time reversal, will enhance and suppress the sign-preserving and the sign-reversing scatterings, respectively [76]. The observation of overall suppression of both intra- and interpocket scatterings in 1-UC FeSe by magnetic vortices [figure 3(c)] violates the QPI selection rules under sign-preserving $s_{++}$-wave pairing. The discrepancy requires a quantitative understanding of the Doppler shift of quasiparticle energy induced by the supercurrent around the vortices, which suppresses the QPI intensities for both magnetic and nonmagnetic scatterings [77]. Alternative explanation for the violation of QPI selection rules is the BZ-folding effect by the glide–mirror-symmetry breaking in 1-UC FeSe. In folded-BZ framework, intra- and interpocket scatterings are symmetry-equivalent and their similar energy dispersions are expected irrespective of the concrete pairing mechanisms. Accordingly, the suppression of intra- and interpocket scatterings by vortices can be naturally reconciled within sign-reversing pairings (refer to table 2).

**Table 2.** QPI selection rules, which are determined from the scattering rate for a transition from the initial state, $(i,\mathbf{k})$, to the final state, $(f,\mathbf{k}')$: $W_{i\to f}(\mathbf{k},\mathbf{k}') \propto \left|u_i(\mathbf{k})u_f^*(\mathbf{k}') \pm v_i(\mathbf{k})v_f^*(\mathbf{k}')\right|^2 \cdot |V(\mathbf{k}'-\mathbf{k})|^2 N_i(\mathbf{k}) N_f(\mathbf{k}')$ ($\pm$ correspond to the scatterings off a potential and a magnetic impurity, respectively). Here, $V$ is the scattering potential, $N_i$ is the partial density of states of band $i$, $v_i(\mathbf{k}) = \text{sign}[\Delta_i(\mathbf{k})]\sqrt{\frac{1}{2}\left[1 - \frac{\varepsilon_i(\mathbf{k})}{E_i(\mathbf{k})}\right]}$ and $u_i(\mathbf{k}) = \sqrt{1-|v_i(\mathbf{k})|^2}$ [76]. "↑" and "↓" denote that the scattering intensity is enhanced and weakened, respectively.

|  | Sign-preserving scattering | Sign-reversing scattering |
|---|---|---|
| Nonmagnetic impurity | ↓ | ↑ |
| Magnetic impurity | ↑ | ↓ |

*Anisotropic $s_{++}$-wave.*—Anisotropic $s_{++}$-wave shows the SC-gap anisotropy, but preserves the sign of SC OP across different Fermi pockets [figure 2(a)] [10,11]. In the full five-orbital tight-binding model, the SOC or the inversion-symmetry breaking lifts the degeneracy of *s*- and *d*-wave SC states and selects *s*-wave as the leading pairing instability [10]. Purely within the random-phase approximation, *d*-wave state is obtained, whereas the *s*-wave state will be obtained if further including the more accurate vertex corrections for the bare Coulomb interactions (*U*-VC) [11]. For the gap structure, earlier ARPES measurements of 1-UC FeSe gave isotropic gap distributions on Fermi surface [8]. Subsequently, high-resolution ARPES corrected the previous results and revealed anisotropic SC gaps [78], whereas the sign structure was not simultaneously determined. Essentially, the gap-function anisotropy originates from $\delta_1/\delta_2$ hybridization and/or the different coupling strengths of involved parameters (e.g., nematic coupling constants [10]) to $d_{xz}/d_{yz}$ and $d_{xy}$ orbitals.

### 3.2 Sign-reversing pairings

*Incipient $s_\pm$-wave.*—The "conventional" $s_\pm$-wave pairing is based on static, band-structure–independent interactions, which predicts no superconductivity upon moving the hole band at $\Gamma$ point below $E_F$. For 1-UC FeSe, while the $\Gamma$ hole band is 80 meV below $E_F$ [8], the conventional $s_\pm$-wave pairing within the electron–hole Fermi-nesting picture [3,4] was theoretically argued via the stabilization by Fermi-surface–based interactions in multiband weak-coupling BCS approximation [13]. In the resulted incipient $s_\pm$-wave pairing, the sign of the SC OP reverses between the incipient hole pockets at $\Gamma$ point and the electron pockets at $M$ points [figure 2(b)] [13-15]. When including the dynamics of excitations allowed by the incipient $\Gamma$ bands near a magnetic instability, the incipient $s_\pm$-wave state contributes significantly to the SC pairing [14]. The induced pairing strength in the incipient bands can be comparable with or even larger than that in the preexisting bands crossing $E_F$ [13].

The high-resolution ARPES revealed the nodeless, moderately anisotropic SC gap in 1-UC FeSe. The gap maxima are along the major axes of the ellipse electron pocket and the gap minima are at $\delta_1/\delta_2$ intersections [figure 4(a)] [78]. The gap-extrema directions are qualitatively described by the gap function of incipient $s_\pm$-wave pairing [figures 4(b) and (c)]. Despite of the large parameter ($\Delta_0 = 112$ meV) in simulation, the calculated result shows the strongest pairing strength only at the center and the corners of unfolded BZ, all of which are Fermi-pocket–free in realistic folded BZ. In ideally nodal situation, the SC gap changes sign at $\delta_1/\delta_2$ intersections where gap nodes occur. In contrast, the hybridization between $\delta_1$ and $\delta_2$ by SOC in practical situation opens a gap at the interpocket crossings and the initial gap nodes evolve into the gap minima. Thus, the detected gap minima at $\delta_1/\delta_2$ intersections can be viewed as a weak indication of the sign reversal on hybridized electron pockets at *M* points.

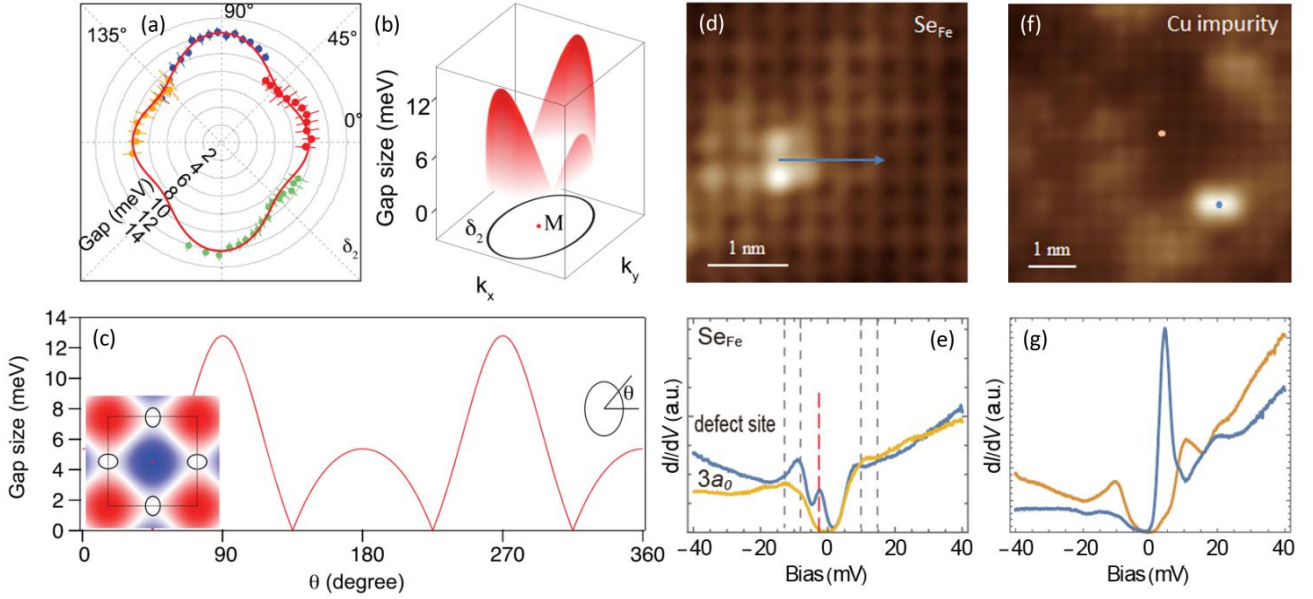

**Figure 4.** SC-gap anisotropy and impurity scatterings in 1-UC FeSe. (a) Anisotropic SC gap on $\delta_2$ pocket. (b,c) Simulation of SC-gap distribution on $\delta_2$ pocket using the gap function, $\Delta = \Delta_0 |(\cos k_x + \cos k_y)/2|$, of incipient $s_\pm$-wave pairing. (d,f) Se$_{Fe}$ antisite defect and Cu impurity, and (e,g) corresponding tunneling spectra. Reproduced from [73,78].

*Extended $s_\pm$-wave.*—In the hole-pocket–absent Fermi-surface topology, the conventional $s_\pm$-wave can be extended alternatively with the sign reversal between hybridized bonding ($\delta_1$) and antibonding ($\delta_2$) electron pockets at *M* points [figure 2(c)] [16,17]. The SOC induces hybridization gap at $\delta_1/\delta_2$ crossings, naturally yielding nodeless SC gaps. Within a 2D model of interacting fermions near (0, π) and (π, 0) (expressed in *unfolded* BZ throughout for clarity), the extended $s_\pm$-wave pairing is favored for larger $\delta_1/\delta_2$ hybridization [17]. The pairing necessarily includes both intra- and interpocket scatterings, where the latter is the leading pairing channel at (π, π), in contrast to that at (π, 0) for conventional $s_\pm$-wave pairing in iron pnictides.

*Odd-parity s-wave.*—Generally, the parity of SC OP is even for spin-singlet pairing. However, in 2-Fe unit cell, there is no parity limitation in each sublattice. Based on symmetry and gauge principles, the possibility of odd-parity *s*-wave spin-singlet SC state was theoretically considered in a low-energy effective model constructed on 2-Fe unit cell for 1-UC FeSe [18]. The analysis concluded that the parity conservation in previous studies of SC states based on the effective *d*-orbital models was violated in fact. The exotic odd-parity *s*-wave pairing shows a real-space sign reversal between the top and the bottom Se layers within 1-UC FeSe. The SC OP in this state is a combination of a normal (***k***,−***k***) *s*-wave pairing between the two sublattices of 2-Fe unit cell and an extended (***k***,−***k***+***Q***) [***Q*** = (π, π)] *s*-wave pairing (η pairing) within the sublattices.

*Nodeless d-wave.*—For smaller $\delta_1/\delta_2$ hybridization, the 2D model of (0, π)−(π, 0) interacting fermions dictates the *d*-wave [17], which is yet nodeless due to the presence of $\delta_1/\delta_2$ hybridization gap as addressed previously [figure 2(d)]. Independently, the spin-fermion description in the effective ***k***·***p*** theory also gives rise to the nodeless *d*-wave state in 1-UC FeSe [19]. This model includes the electronic coupling to the fluctuations of checkerboard

AFM order, which well produces the band structure and gap anisotropy of 1-UC FeSe seen in ARPES.

The magnetically selective impurity effects on electron pairing suggest that the nonmagnetic scatterings are especially important in determining the pairing symmetry (see table 1). Naturally, Se is purely nonmagnetic. By DC magnetization measurements, Cu dopants have been verified as nonmagnetic impurities as well [79]. Distinct from the scattering results of Zn, Ag, K adatoms [70] and Fe vacancy [73] [figures 3(d1)–(e2)], the scatterings off nonmagnetic $Se_{Fe}$-antisite defect and Cu impurity in 1-UC FeSe induce sharp in-gap bound states [figures 4(d)–(g)], which violates the Anderson's theorem [72]. In sign-reversing multiband pairing, the dominated interpocket scatterings by local strong impurities see the sign reversal of SC-gap function and can induce the intragap states [79]. Presumably, the detection of the bound states on nonmagnetic defects in 1-UC FeSe is supportive of the sign-reversing pairing.

The discrepancy between different impurity scatterings with diverging pairing conclusions [figures 3(d1)–(e2) vs. figures 4(d)–(g)] mainly originates from the experimental uncertainty of impurity magnetism and/or bound-state occurrence. As mentioned earlier, the identification of nonmagnetic impurities is generally challenging in experiments. The perceived nonmagnetic impurities may consist of the component of magnetic moment, whereas the nominally magnetic impurities may behave magnetically neutral. Besides the complexities of the magnetic properties of impurities, pinning down the existence of bound states induced by nonmagnetic scatterers has proven nontrivial. Especially for weak scattering potentials near Born limit, the bound states are virtually invisible. Both well-defined strong nonmagnetic scatterers and clear identifications of impurity bound states are highly desired for further experimental investigations.

### 3.3 Pairing interactions

The pairing interactions "gluing" Cooper pairs vary in different pairing proposals for 1-UC FeSe. The physically concrete pairing interactions involve EPC and exchange of quantum-order fluctuations, including spin, nematic and orbital fluctuations. There are other pairing forces considered in theories, like multiorbital Hubbard interactions [53], which are quite theoretically technical and will not be further discussed here.

*EPC*.—EPC is responsible for $s_{++}$-wave pairing, which was originally established in the BCS theory. The "replicas" of the main bands near $\Gamma$ and *M* points [figures 5(a) and (b)] were detected by ARPES in 1-UC FeSe on different substrates ubiquitously, including $SrTiO_3$(110) [80] and rutile $TiO_2$(100) [36] besides $SrTiO_3$(001) [47]. Model calculations qualitatively sketch the replica bands approximately 100 meV below the main bands [figure 5(d)], where both electron and hole bands are assumed to couple to a dispersionless 80-meV phonon. Provided the reported phononic mode with comparable energy in a previous study, the replica bands were interpreted as the coupling of FeSe electrons with 100-meV oxygen optical phonon in $SrTiO_3$ with a small momentum transfer *q*.

Ultraviolet photoemission spectroscopy shows gradually enhanced spectral weight of Ti−O-bonding–derived peak in 1-UC FeSe as superconductivity is strengthened with consecutive annealing [figure 5(e)] [58]. Reminiscent of the replica-band–typified optical phonon that propagates along the Ti−O direction, the enhanced Ti−O bonding is suggestive of the simultaneously enhanced EPC across FeSe/$SrTiO_3$ interface. The phononic signatures (replica bands and Ti−O bonding) detected by different photoemission spectroscopies highlight the importance of EPC [81,82] in superconductivity of 1-UC FeSe.

The phononic-coupling interpretation of the replica bands across FeSe/$SrTiO_3$ interface via shaking off the bosonic quanta [47] was challenged by the observation of replica bands in bare $SrTiO_3$ [80,83,84]. At semi-quantitative level, by combining local density approximation (LDA) and dynamical mean-field theory, the replica bands in 1-UC FeSe can be also understood as the calculated Fe-$3d_{3z^2-r^2}/3d_{xy}$ ($\Gamma$/*M*) bands renormalized by the electron correlations [85]. From this perspective, the replicas are purely of electronic nature without reference to the interactions with optical phonons in SrTiO$_3$. More quantitatively, the semiclassical dielectric theory precisely reproduced the replica-band features in 1-UC FeSe without fitting parameters [86]. In this scenario, the replica bands alternatively result from the extrinsic energy loss of photoelectrons by exciting SrTiO$_3$ Fuchs–Kliewer surface phonons in the photoemission processes. Besides the theoretical challenges, the detected phononic signatures persisting above $T_c$ [figure 5(c)] contradict the intimate correlation between the pair-binding EPC and superconductivity. Actually, according to the phenomenological theory in Ref. [22] describing the replica bands, small-$q$ EPC with large associated Bohm effective charge (e.g., −8 for 100-meV SrTiO$_3$ optical phonon) only serves to enhance the sign-reversing pairing by (π, π) spin fluctuations regardless of the pairing symmetry. While the EPC cannot be fully excluded as the pairing glue for SC condensation, the replica anomaly beyond SC region at least implies that there exists the phononic component uninvolved in electron pairing in 1-UC FeSe.

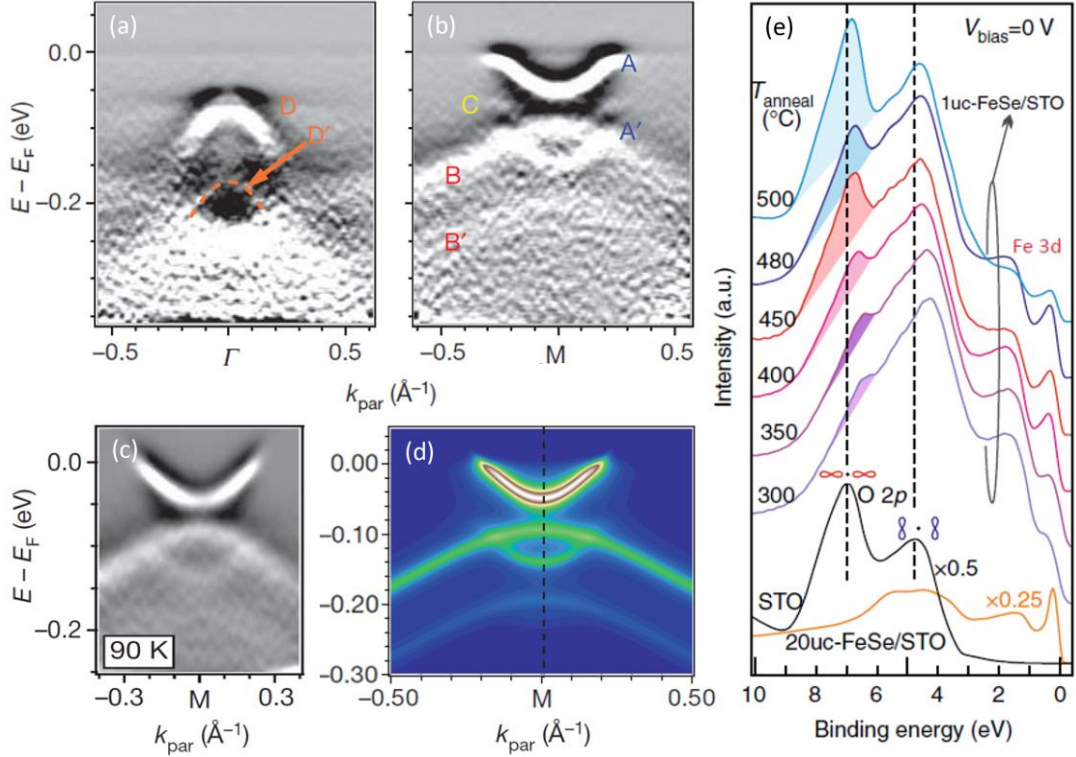

**Figure 5.** Phononic signatures in 1-UC FeSe. (a,b) Band structures near Γ and $M$ points at 16 K, where the replica bands are denoted as A', B' and D'. (c) Replica band near $M$ point at 90 K. (d) Spectral-function calculation with both hole and electron bands coupling to a dispersionless 80-meV phonon. (e) Valence spectra of O-2$p$- and Fe-3$d$-orbital–derived features, where the Ti−O-bonding–related peaks are highlighted by shaded regions. Reproduced from [47,58].

*Spin fluctuations.*—According to the BCS theory [55],

$$\Delta_k = -\sum_l \frac{V_{kl}\Delta_l}{2E_l}, \qquad (5)$$

the repulsive interaction potential ($V_{kl} > 0$) corresponds to the sign reversal of SC gap [sgn($\Delta_k$) ≠ sgn($\Delta_l$)]. As the repulsive exchange interaction, spin fluctuations have been established as the pair-binding glue in sign-reversing pairing. In multiband $s_\pm$-wave pairing of iron pnictides, spin fluctuations as the collective bosonic excitations originate from the AFM ground state. In 1-UC FeSe, various AFM orders of the Fe layer have been proposed as the magnetic ground states by first-principles calculations of the band structures, including collinear, block-checkerboard, pair-checkerboard and checkerboard AFM configurations [figures 6(a)–(d)] [38-41]. Among these proposals, the calculated band dispersions based on the checkerboard AFM order best match the ARPES results [41]. Note that the double-TiO$_x$ termination of SrTiO$_3$ has not been considered throughout all the calculations. It will be interesting to see how the AFM orderings are modified as the double-TiO$_x$ effect is included.

The long-range AFM order was revealed in the parent phase of 1-UC FeSe by magnetic-hysteresis measurements [42]. In ferromagnet–antiferromagnet heterostructure, the magnetic-exchange bias effect (MEBE) occurs distinctly typified by the shifting of magnetic-hysteresis loop with its center deviating from $H = 0$ point. Consistently, in Fe$_{21}$Ni$_{79}$/as-grown 1-UC FeSe heterojunction [figure 6(e)], where Fe$_{21}$Ni$_{79}$ is the reference ferromagnetic (FM) layer, the center of the magnetic-hysteresis loop shifts away from $H = 0$ [figures 6(f) and (g)]. Additionally, when the cooling magnetic field is reversed, the shifting direction accordingly reverses [figures 6(f) and (g)]. Both the phenomena indicate the existence of MEBE, which supports the antiferromagnetic order in the electron-undoped 1-UC FeSe.

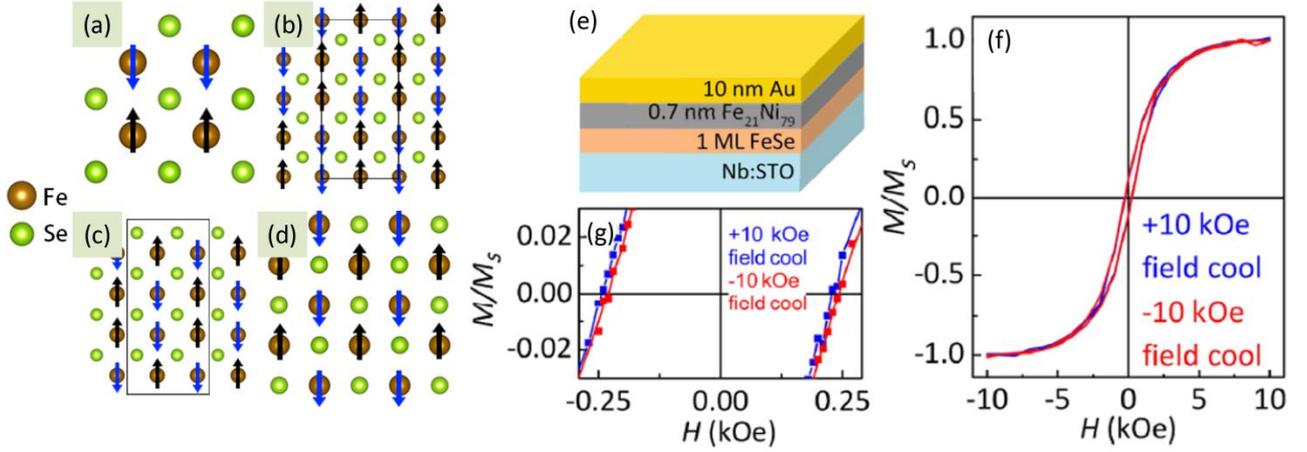

**Figure 6.** AFM orders in 1-UC FeSe. (a–d) Theoretically proposed collinear [38], block-checkerboard [39], pair-checkerboard [40] and chequerboard AFM orders [41]. (e) Schematic of Fe$_{21}$Ni$_{79}$/as-grown 1-UC FeSe heterostructure. (f,g) Magnetic-hysteresis loops of the heterojunction in (e) at 5 K. Reproduced from [42].

The AFM order within the FeSe layer can physically maintain the AFM spin fluctuations upon heavy electron doping. In multiband sign-reversing pairing, Cooper-paired electrons scattered between $\Gamma$ hole pockets and phase-inversed $M$ electron pockets are predicted to interact by AFM spin fluctuations. When the hole pockets at $\Gamma$ point are absent as in 1-UC FeSe, the paired electrons linked by AFM spin fluctuations are alternatively scattered mainly between the electron pockets at adjacent $M$ points with sign reversal. The electronic coupling to AFM spin fluctuations will reconstruct the quasiparticle density-of-states spectrum with an additional hump outside the coherence peak, which has been possibly captured by a recent STS study of 1-UC FeSe [48].

In frustrated Heisenberg AFM model,

$$H = J_1 \sum_{l,\langle i,j \rangle} \mathbf{S}_{il} \cdot \mathbf{S}_{jl} + J_2 \sum_{l,\langle\langle i,j \rangle\rangle} \mathbf{S}_{il} \cdot \mathbf{S}_{jl}, \tag{6}$$

the nearest- ($J_1$) and the next-nearest-neighbor AFM exchange couplings ($J_2$) are the main physical parameters in tuning the phase boundaries. Empirically, $J_1$ and $J_2$ favor ($\pi$, $\pi$) and ($\pi$, 0) AFM interpocket scatterings, respectively. For iron pnitides with $\Gamma$ hole pockets, the spin-fluctuation–mediated SC state is mainly understood in terms of $J_2$-dominated ($\pi$, 0) pairing [87]. For 1-UC FeSe, there have been more detailed studies beyond the above-mentioned empirical pairing criteria by focusing on $J_1$- and $J_2$-type AFM fluctuations mimicking $J_1$ and $J_2$ AFM exchange interactions [50]. The results favor $J_2$-type spin fluctuations in $s$-wave pairing, whereas $J_1$-type spin fluctuations in nodeless $d$-wave case. Provided the controversy of pairing symmetry in 1-UC FeSe to date, whether $J_1$ or $J_2$ precisely dominates the electron pairing therein remains to be explored.

As mentioned in part 2, GGA+$U$ calculations show that the collinear AFM phase of FeSe layer is stabilized by the tensile strain in 1-UC FeSe [65]. However, the $T_c$ facilitated by the tensile strain with at a magnitude of only ~10 K [figure 1(d)] is comparatively low in the context of AFM-spin-fluctuation–mediated superconductivity. Hence, either collinear AFM phase or spin-fluctuation pairing in 1-UC FeSe is still questionable, which requires detailed theoretical explorations in future.

*Nematic fluctuations.*—The enhancement of superconductivity was addressed by several works in the vicinity of nematic transition by nematic fluctuations [88,89]. The nematic order exists in multilayer FeSe films on SrTiO$_3$ [90], but is absent in FeSe at 1-UC limit [91]. The proximity to a nematic phase in $T$–thickness diagram suggests the possibility of nematic fluctuations in 1-UC FeSe. Based on a sign-problem–free model using quantum Monte Carlo, calculations reveal the simultaneously enhanced nematic fluctuations and superconductivity in 1-UC FeSe at intermediate couplings [49]. For the detailed pairing channels, the nematic fluctuations all enhance the SC order [50]. These theoretical results establish the potential intimate correlation between nematic fluctuations and the enhancement of superconductivity for 1-UC FeSe.

*Orbital fluctuations.*—More sophisticated calculations proposed the orbital-fluctuation–mediated pairing in HEDISs by considering the higher-order many-body effects [11], including $U$-VC, which is beyond the random-phase approximation. The orbital fluctuations consist of ferro- and antiferro-orbital fluctuations, which are realized by the complex interplay of Coulomb and electron–phonon interactions. Moderate orbital fluctuations yield strong attractive pairing interactions and correspond to $s_{++}$-wave pairing [12]. The concrete signals responsible for the orbital fluctuations remain unaddressed yet and are highly desired for further theoretical input.

*Cooperative-pairing conjecture.*—At FeSe/SrTiO$_3$ interface, a 2D charge system mainly induced by oxygen vacancies exists in the form of both itinerant and immobile electrons. The electron-density fluctuations and the Coulomb interactions in the interfacial charges screen the cross-interface EPC [92,93]. Besides the partial phononic component unrelated to SC pairing (section "*EPC*" in Part 3.3), the effective EPC constant, $\lambda(x)$, will be reduced, rendering that the EPC alone cannot explain the high $T_c$ in 1-UC FeSe [92,94]. Even in the limiting situation without EPC screening, DFT calculations found the phonon-mediated contribution to $T_c$ is only a small fraction, which is one order of magnitude lower than that typically observed in experiments [95]. Given the experimental unavailability of both nematic and orbital fluctuations, the cooperative pairing involving multiple channels [13], including both EPC and spin fluctuations, is the most plausible scenario responsible for the high-temperature superconductivity in 1-UC FeSe.

## 4. Universal SC mechanism

The 1-UC FeSe is strongly electron-correlated predominately dependent of Hund's interaction [45] and is consequently driven into an orbital-selective Mott phase [46]. Upon heavy electron doping, the highly enhanced interface superconductivity appears near spin-density-wave (SDW) [25] or AFM instability [42]. The remarkable resemblance of 1-UC FeSe to previous unconventional superconductors in phenomenology probably implies a common SC mechanism of high-temperature superconductivity among different compounds.

### 4.1 Electron correlation and orbital-selective Mott phase

While iron pnictides are moderately electron-correlated, where the observed band dispersions are largely consistent with the LDA predictions, the 1-UC FeSe exhibits strong electron correlations [45,46]. Specifically, for 1-UC FeSe, compared with LDA calculations [figure 7(b)], the $d_{xy}$ band is more prominently renormalized than $d_{xz}/d_{yz}$ bands as revealed by ARPES, showing a nearly flat hole-like dispersion [figure 7(a)]. The strongly orbital-selective band renormalization, together with the non-rigid band shift as electron doping [44], manifests strong electron correlations in 1-UC FeSe. Detailed theoretical study combining both DFT and dynamical mean-field theory used the oxygen-vacant phase of 1-UC FeSe/SrTiO$_3$ to simulate the interfacial FeSe/SrTiO$_3$ system in experiments [45]. Controlled by the Se−Fe−Se angle, the electron correlations were found to be predominately dependent of Hund's interaction. As comparison, 1-UC FeSe/SrTiO$_3$ without oxygen vacancies shows weaker electron correlations, suggesting electron doping makes the system more correlated. Furthermore, the tensile strain in 1-UC FeSe film from the SrTiO$_3$ substrate also serves to enhance the electron correlations [33].

With increasing temperature, the $d_{xy}$ orbital of 1-UC FeSe completely loses spectral weight, whereas the $d_{yz}$ orbital remains metallic [figures 7(c)–(e)]. By slave-spin mean-field theory [96], the orbital selectivity of the electron-localization behavior as raising temperature can be understood in proximity to an orbital-selective Mott phase. With sufficiently strong intraorbital Coulomb repulsion $U$ and Hund's coupling $J$, the orbital-selective Mott phase dictates that the $d_{xy}$ orbital be orbital-selectively Mott-localized and other orbitals keep itinerant, both well describing the orbital-dependent observations in 1-UC FeSe. Physically, the orbital selectivity originates from the different electron repulsions and relative bandwidths of Fe-3$d$ orbitals. The orbital-selective Mott phase in 1-UC FeSe is intriguingly different from the fully Mott-insulating state in cuprates.

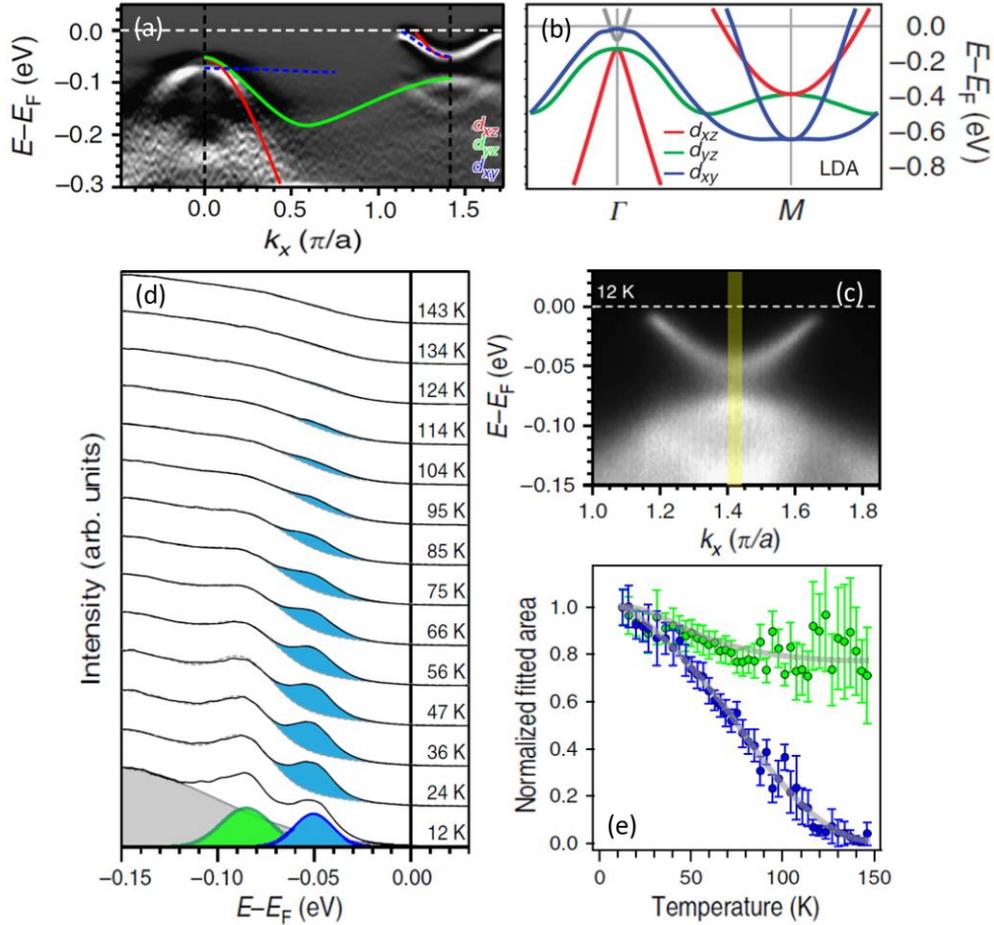

**Figure 7.** Electron correlation and orbital-selective Mott phase in 1-UC FeSe. (a) Band spectra along Γ-$M$ direction. (b) Schematic band structures by LDA calculations [97]. (c) Band spectra near $M$ point. (d,e) Temperature dependence of $d_{xy}$ (blue) and $d_{yz}$ (green) spectral weights extracted within the yellow-shaded region in (c). Reproduced from [46].

### 4.2 Comparison with other HEDISs

Besides 1-UC FeSe/SrTiO$_3$, the HEDISs also include $A_x$Fe$_{2−y}$Se$_2$ ($A$ = K, Rb, Cs, Tl) [98] and (Li$_{1−x}$Fe$_x$)OHFe$_{1−y}$Se [99]. All of them share nearly the same band structures, Fermi-surface structure and SC-gap (Δ) symmetry [5,7,8,30]. Thus, the experimental results obtained in $A_x$Fe$_{2−y}$Se$_2$ and (Li$_{1−x}$Fe$_x$)OHFe$_{1−y}$Se can serve as references for 1-UC FeSe.

*Nodeless SC gap.*—Similar to 1-UC FeSe, ARPES experiments also show nearly isotropic, nodeless SC gaps for $A_x$Fe$_{2-y}$Se$_2$ and (Li$_{1-x}$Fe$_x$)OHFe$_{1-y}$Se at electron pockets at $M$ points ($k_z = 0$) [5,7]. In the unfolded-BZ framework, or when the band-folding potential is weak, there is no sign reversal and presumably, no gap nodes, around individual $M$ points in multiband $d$-wave pairing case. Thus, the detected nodeless SC gaps at $M$ points cannot be solely regarded as the evidence for $s$-wave pairing. However, the gap nodes do appear along (0, 0)−(π, π) direction. Since there are no Fermi pockets at Γ point at $k_z = 0$ for HEDISs, the gap structure on $κ$ electron pocket at BZ center at $k_z = π$ will be crucial in differentiating $s$- and $d$-wave pairings. Experimentally, in K$_x$Fe$_{2-y}$Se$_2$, the ARPES results found that $κ$ pocket shows nodeless gap distribution as well, which favors $s$-wave pairing instead of nodeless $d$-wave [100], although $s_{++}$- and extended $s_{\pm}$-wave cannot be further distinguished.

*Sign-reversing pairing.*—For both $A_x$Fe$_{2-y}$Se$_2$ and (Li$_{1-x}$Fe$_x$)OHFe$_{1-y}$Se, inelastic neutron scatterings revealed magnetic resonant modes in spin-excitation spectra in the SC states [101-104]. The resonant modes demonstrate an OP-like behavior with an onset temperature at $T_c$, highlighting their intimate correlations with superconductivity. Especially, the modes generally show an energy scale ($E_{res}$) upper-bounded by 2Δ and were observed near the nesting vectors of Fermi pockets at adjacent $M$ points. These observations fall within the universal scaling relation ($E_{res}$ ~0.64 × 2Δ) of magnetic resonances summarized from a large number of unconventional cuprates and iron pnictides [105], and coincide with the spin-susceptibility calculations in extended $s_{\pm}$- [106] or nodeless $d$-wave pairing [107]. Thus, the neutron experiments support the sign reversal of SC OP in HEDISs.

Theoretically, the magnetic resonance modes appear as the resonance peaks in the imaginary parts of dynamical spin susceptibilities, $χ''(Q,ω)$, and are energy-limited below 2Δ due to the excitations of particle–hole pairs. The coherence factor characterizing the process of particle–hole excitation is expressed as

$$1 - \frac{\varepsilon_q \varepsilon_{q+Q} + \Delta_q \Delta_{q+Q}}{E_q E_{q+Q}}, \tag{7}$$

where $\varepsilon_q$ is the single-particle dispersion, $\Delta_q$ is the gap function and $E_q$ is the quasiparticle energy [108]. It can be seen that the coherence factor remains a finite value only for the gap functions at the momenta of the excited particle–hole pairs, $\Delta_q = -\Delta_{q+Q}$, with reversed signs. This explains in principle that the emergence of magnetic resonance modes within 2Δ near Fermi-nesting vectors $Q = (π, π)$ is experimentally taken as the evidence for sign-reversing pairing in HEDISs. Independently, by the recently proposed Hirschfeld–Altenfeld–Eremin–Mazin (HAEM) method [109], the QPI intensity of intrapocket scattering, $q_1$, for Zn-doped (Li$_{1-x}$Fe$_x$)OHFe$_{1-y}$Zn$_y$Se was quantified as a function of bias voltage [110]. The obtained dispersion well matches the $s_{\pm}$-wave pairing, revealing the sign-reversing scenario as concluded from the neutron-scattering results.

The 1-UC FeSe is easily degraded upon exposure in atmosphere for *ex situ* measurements and its signals would be merged into the background of far thicker protection layer or substrate. Because of these difficulties, the neutron-scattering results, which are crucial in determining the sign structure of gap function via magnetic modes, are unavailable yet for 1-UC FeSe. In view of the similarities among different HEDIS compounds, the signatures of sign reversal observed in other HEDISs are instructive for the future understanding of the gap structure of 1-UC FeSe.

*Magnetic order and spin fluctuations.*—In multiband sign-reversing pairing, the sign reversal is seen predominantly by interpocket scatterings at Fermi-nesting vectors, where the AFM spin fluctuations occur and mediate the high-temperature superconductivity. This is made possible by the AFM orders in HEDISs, which are established by extensive theoretical and experimental studies. In detail, for (Li$_{1-x}$Fe$_x$)OHFe$_{1-y}$Se, (Li,Fe)(OH) and FeSe layers were proposed to be FM [111] and AFM [112,113], respectively. In the FeSe layer, while the striped AFM order is energetically favored [112], the checkerboard AFM order yields the calculated band structures consistent with ARPES results [113]. Experimental signatures of the suppressed specific-heat jump by magnetic field [99] and the two-magnon excitations in Raman spectra [114] suggest the AFM nature of SC and insulating states of $(Li_{1-x}Fe_x)OHFe_{1-y}Se$, respectively. Nevertheless, the detailed spin structures in AFM phase cannot be precisely distinguished. In $(Li_{1-x}Fe_x)OH(Fe_{1-y}Li_y)Se$, isothermal-magnetization measurements show magnetic-hysteresis loops characteristic of FM orders instead coexisting with superconductivity [115]. Unlike 1-UC FeSe case with purely AFM signal, the experimental phenomena regarding magnetic order in $(Li_{1-x}Fe_x)OHFe_{1-y}Se$ are complicated without a unifying understanding due to the uncertainty of exact chemical constituents in different literature. For $A_xFe_{2-y}Se_2$, AFM and SC regions are phase-separated at mesoscopic scale in space [116]. The AFM ground states of HEDISs are physically responsible for the spin fluctuations in sign-reversing pairing. Despite the details in individual HEDIS compounds, the AFM phases universally exist therein in different forms as the ground states, reminiscent of those addressed in 1-UC FeSe (section "*spin fluctuations*" in Part 3.3).

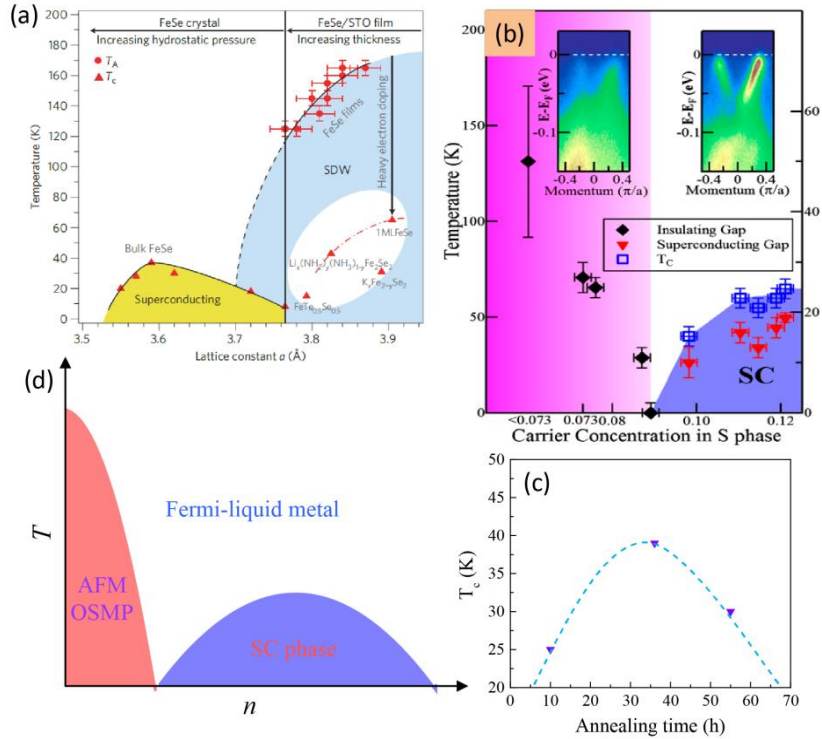

**Figure 8.** Phase diagram for FeSe. (a) $T_c$–$a$ diagram for bulk and 1-UC FeSe. (b) Temperature–carrier concentration diagram for 1-UC FeSe. (c) Annealing-dependent $T_c$ of 5-UC FeSe by electrical transport [26], qualitatively sketching the $T_c$–carrier density diagram for 1-UC FeSe because the conducting channel in 5-UC FeSe is only from the interfacial FeSe layer. (d) Schematic $T_c$–$n$ diagram for 1-UC FeSe based on [26,42,43,46]. $n$, carrier density; OSMP, orbital-selective Mott phase. Reproduced from [25,43].

*Electron correlations and Mott physics.*—The SC states of HEDISs have been theoretically proposed on the verge of doped Mott insulators in phase diagram. Using two-orbital model and slave-rotor method, the ordered Fe vacancies in HEDISs have been shown to enhance the electron correlations, and thus the tendency towards Mott transition [117]. As an example, the insulating parent of $(K,Tl)_yFe_xSe_2$ superconductor [$(K,Tl)_yFe_{1.5}Se_2$] can be a Mott insulator at ambient pressure [117]. Electron doping by partial filling of the Fe vacancies therein drives the Mott transition to an orbital-selective non-Fermi-liquid metal, implying superconductivity arises possibly as a doped Mott-insulating state [118,119]. Correspondingly, in the experiments of $K_xFe_{2-y}Se_2$, the SC phases exhibit large-scale spectral-weight transfer to the insulating phases, resembling the opening of Mott–Hubbard gap in the insulating states [120]. Furthermore, reminiscent of the 1-UC FeSe results, $A_x$Fe$_{2-y}$Se$_2$ ($A$ = K, Rb) shows heavily renormalized $d_{xy}$ band in experiments compared to LDA, indicating strong electron correlations [121]. Increasing temperature further drives the system from a metallic state to a selectively $d_{xy}$-orbital–localized state, where other orbitals remain itinerant. As the situation in 1-UC FeSe, these phenomena are consistently understood as an orbital-selective Mott phase with sufficiently strong on-site Coulomb interaction and Hund's coupling, which is different from but in close proximity to the fully Mott-insulating phase.

### 4.3 Phase diagram

The SDW selectively exists in multilayer FeSe films and is strengthened with reduced thickness or equivalently, increased tensile strain [25]. As superconductivity in 1-UC FeSe is induced upon the cross-interface charge doping, the otherwise pronounced SDW is suppressed "anomalously". This means that the SC states of 1-UC FeSe are located near the quantum criticality of SDW instability [figure 8(a)]. The competition between superconductivity and SDW follows the spirits of cuprates and iron pnictides, where the "dome"-like unconventional SC order emerges by suppressing AFM Mott phase via charge doping, except that the control parameter in the case of 1-UC FeSe here is replaced by the film thickness or the tensile strain. More essentially, the presence of strong electron correlations and orbital-selective Mott phase suggests 1-UC FeSe is near the phase boundary of the Mott-insulating state [46]. Upon electron doping, the AFM order "fingerprinted" by MEBE disappears [42] and superconductivity appears as dome-like in the $T_c$–electron doping phase diagram [26,43] [figures 8(b)–(d)]. Similarly, in the phase diagram of (Li$_{1-x}$Fe$_x$)OHFe$_{1-y}$Se, AFM-type SDW insulating state is in proximity to the dome-shaped superconductivity, which was suggested to originate from the weak AFM fluctuations [122]. Further considering the similarities summarized earlier between 1-UC FeSe and other HEDISs, all these results tentatively establish a unifying understanding of high-$T_c$ superconductivity in cuprates, iron pnictides and HEDISs.

## 5. Topological states

Topological matters show Dirac-cone–like band structures with linear dispersions described by topologically nontrivial invariants. Theoretical explorations successively revealed various topological phases potentially existing in 1-UC FeSe from different perspectives, including topological-metallic, topological-insulating [51], 2D Dirac-semimetallic [44], AFM quantum spin Hall (QSH) [41] and quantum anomalous Hall (QAH) phases [52]. The combination of high $T_c$ and topological phases possibly endows 1-UC FeSe/SrTiO$_3$ with high-temperature topological superconductivity [53], which supports MBSs at the zero-dimensional topological edges, like magnetic vortices.

### 5.1 Topological-metallic and topological-insulating phases

Topological-metallic and topological-insulating phases were predicted in 1-UC FeSe under the cooperative effects of substrate tensile-strain field and SOC [51]. The tensile strain exerted in 1-UC FeSe leads to the lattice distortion and the change of electron hoppings. By incorporating interorbital hopping, a gap is opened at $M$ point and the hole band is suppressed at $\Gamma$ point, resulting in the band structures well comparable to the ARPES experiments. Additionally adding SOC, the band gap at $M$ point is driven to undergo a closing–reopening process, indicating a topological phase transition. The gapless edge states emerging following the gap reopening clearly exhibit the signatures of Dirac-type topological phases, which are topologically metallic or insulating depending on the relative position of the hole-band top at $\Gamma$ point and the electron-band bottom at $M$ point. However, with even (two) nontrivial Dirac-cone edge dispersions in 2-Fe unit cell, the topological phases are unstable against any perturbations that break the nonsymmorphic lattice symmetry. When including the substrate Hamiltonian, the system with weak topologies can be stabilized and driven into strong topological phases with only one nontrivial

Dirac cone survived. Thus, both the substrate coupling and the SOC are necessary for the robust topological phases in 1-UC FeSe. Provided that the SOC is stronger than the opened gap at *M* point by lattice distortion, the robust topological phases can always be achieved.

### 5.2 2D Dirac-semimetallic phase

While the annealed SC 1-UC FeSe exhibits the band structures without Dirac cones, its electron-undoped parent phase was demonstrated to be the 2D Dirac semimetal [44]. The Dirac-like linearly dispersive bands are located at $k_x \sim 0.8\pi/a$ ($\Lambda$) between $\Gamma$ and *M* points [figures 9(a)–(c)]. The Fermi level precisely crosses the Dirac point without touching any other bands in the BZ [figure 9(a)], indicating the semimetallic nature of the parent 1-UC FeSe. In multilayer FeSe films, the Dirac-cone dispersions coexist with the nematicity and accordingly disappear above the nematic transition temperature [123]. The observed Dirac bands in nematicity-absent parent 1-UC FeSe contradict the multilayer situation, signifying the different roles of electron doping in FeSe films with varied thickness.

### 5.3 AFM QSH state and topological edge states

First-principles calculations reveal the experimentally self-consistent band structures for 1-UC FeSe based on the checkerboard AFM order with SOC effect [41]. From this point, the spin Chern number obtained by integrating the spin Berry curvature gives an integer value of −1. Correspondingly, the spin Hall conductance calculated by Kubo formula is quantized within the energy window of the SOC-induced gap [figure 9(d)], demonstrating an AFM QSH state. Different from the conventional QSH state, the AFM QSH state breaks the time-reversal symmetry, but alternatively preserves the combined symmetry of time-reversal and primitive lattice-translation.

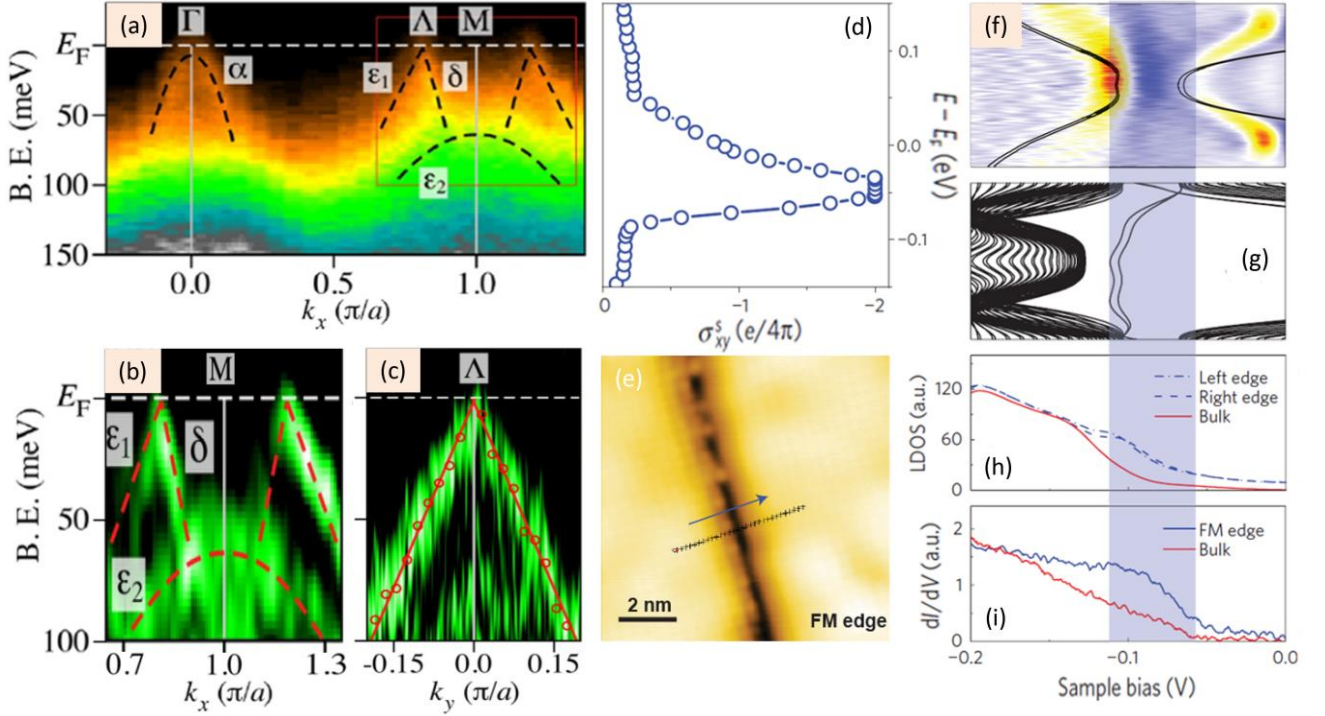

**Figure 9.** Topological phases in 1-UC FeSe. (a) Band spectra along $\Gamma$-*M* direction of undoped 1-UC FeSe and (b,c) corresponding close-up views. (d) Calculated energy dependence of spin Hall conductance. (e) Topographic image of FM edge along [100]-orientated grain boundary. (f) Band structures around *M* point. (g) Calculated band structures of FeSe/SrTiO$_3$ ribbon with FM edges. (h,i) LDOS simulations and tunneling spectra of FM edge and bulk states. Reproduced from [41,44].

The AFM QSH state yields the edge states of topological nature. By Wannier-function calculations, the 1D FM

edge of FeSe/SrTiO$_3$ ribbon directly shows the Dirac-type edge states within the SOC gap [figures 9(f) and (g)]. In the same gap-energy range, the FM edge state around −0.1 eV was detected with enhanced spectral intensity by tunneling spectrum at the [100]-orientated grain boundary [figures 9(e) and (i)]. Theoretically, the spectral lineshape of the observed FM edge state was reproduced by local-density-of-states (LDOS) simulations [figure 9(h)]. The results are similar in the AFM-edge case. These findings in 1-UC FeSe are consistent with the spectroscopic expectation of the AFM QSH edge state. Besides the rough energy consistency for the edge mode between experiment and theory, further experimental evidence is still required to fully pin down the existence of topological edge states.

### 5.4 QAH state

Based on first-principles tight-binding calculations, the spin-tunable QAH state can be induced in 1-UC FeSe from the AFM QSH state by a time-dependent irradiation of left- or right-handed circularly polarized light [52]. The spin-up and spin-down bands respond oppositely to the circularly polarized light of opposite handedness. As increasing the light intensity, the spin degeneracy of the band dispersions is lifted at *M* point. While the gap of the spin-up band increases monotonically, the gap of the spin-down band closes and reopens. The band-gap closing–reopening process suggests the occurrence of a band inversion (BI), indicative of a topological phase transition. Different from the conventional BI (type-I BI), the above BI (type-II BI) is triggered by the light-induced, handedness-sensitive effective SOC, which inverts the band an even number of times and breaks the time-reversal symmetry. The spin-down band experiencing the type-II BI is accompanied by a sign change of the Chern number. Hence, the system is topologically nontrivial for both spin-up and spin-down components, rendering a high-Chern-number (±2) QAH state [52].

The light-induced QSH-to-QAH phase transition in 1-UC FeSe is concretely identified by the calculations of both the edge states of FM-edged FeSe ribbon and the quantized Hall conductivity. In the edge-state simulations after the type-II BI, the left- and right-edge–projected band structures show that each edge consists of two spin-opposite edge states sharing the same propagating direction as expected in QAH phase. Independently, within the energy window of the band gap for both handednesses of light, the Hall conductivity is quantized at −2e$^2$/h and 2e$^2$/h for the left- and right-handed circularly polarized light, respectively, also precisely within the QAH scenario.

### 5.5 Topological-SC phases

From the view point of group theory and topological theory, the electron-doped 1-UC FeSe possibly supports topological-SC phases in the spin-triplet, orbital-singlet *s*-wave pairing channel [53]. The point group of 1-UC FeSe is $D_{4h}$. The pairing symmetries of the SC OPs are classified by the irreducible representations of $D_{4h}$ group. Physically, both (***k***,−***k***) and (***k***,−***k***+***Q***) Cooper pairings are symmetry-allowed. Only (***k***,−***k***) pairing channel is considered to avoid orbital-parity mixing and inversion-symmetry breaking. To find out the possible pairing channels supporting the topological superconductivity, several constraints are imposed to all the (***k***,−***k***) pairings, including being nodeless-gapped, having odd parity and being C4-symmetric. The odd-parity pairing states of $E_u^{(1)}$, $E_u^{(2)}$, $E_u^{(3)}$, $A_{1u}^{(1)}$ ($\Delta_1$–$\Delta_4$) eventually survive, all demonstrating topological-SC states dictated by the bulk–boundary correspondence. The Bogoliubov–de Gennes Hamiltonian describing the SC states gives the edge spectra explicitly supporting the Andreev bound states characteristic of topological superconductivity. Quantitatively, the bulk topological properties for SC states in $\Delta_1$–$\Delta_4$ channels are described by topologically nontrivial invariant. Thus, the topological-SC states can emerge in the nodeless odd-parity pairing channels.

## 6. Perspectives

While the dramatic $T_c$ enhancement in 1-UC FeSe can be semi-quantitatively interpreted by electron doping and tensile strain, the essential pairing mechanism remains unresolved. Note that the impurity states evolve systematically as finely tuning the scattering potentials [124]. Thus, in future investigations, one of the directions for addressing the pairing problem in 1-UC FeSe can be identifying the nonmagnetic-impurity bound states by fine tuning of the scattering strength. Additionally, different pairings mainly differ in the SC-gap functions and the sign structures. For 1-UC FeSe, the ellipticity of Fermi pockets is small and the hybridized $\delta_1/\delta_2$ pockets are hardly resolved by generic spectroscopies. To get deeper insights into the pairing scenarios, the ultrahigh-resolution QPI technique [125] is required for sketching the fine structures of intrapocket scatterings in QPI patterns and extracting the SC-gap distributions on Fermi pockets. These are the prerequisites for further mapping the sign structures over the whole Fermi surface according to the QPI selection rules and determining the SC-gap functions by comparing with proposed pairings.

HAEM [109] and phase-referenced methods [126] are the recently proposed approaches for QPI analyses. For HAEM, the principle idea is to trace the bias dependence of the quantified scattering-vector intensity and compare it with the theoretical simulations by different pairings. For the phase-referenced QPI, both magnitude and phase information of the QPI patterns are reserved, in sharp contrast to the conventional QPI techniques that only keep the magnitude information. The principle idea is that, for FFT-QPI at defect-bound-state energies, $E$ and $-E$, the scatterings nesting sign-reversing (sign-preserving) Fermi sheets yield sign-opposite (sign-equal) *complex* QPI signals. Both these recently developed QPI methods have been applied to determine the $s_\pm$-wave pairing in the bulk materials of FeSe [125], $(Li_{1-x}Fe_x)OHFe_{1-y}Zn_ySe$ [110] and LiFeAs [126], whose applications to 1-UC FeSe will be crucial for the pairing understanding therein. Furthermore, the ARPES-extracted real and imaginary parts of the self-energy involve the information about the interaction of electrons with bosonic modes, like magnons, phonons, and excitons [127]. Extracting the self-energy in 1-UC FeSe by high-resolution ARPES will also help clarify the debate regarding the pairing interactions therein.

Finally, the newly arisen directions for pursuing the MBSs turn out to be the iron-based high-temperature superconductors. In recent ARPES experiments, the Dirac-type spin-helical surface state and $s$-wave SC gap were simultaneously observed in the bulk $FeTe_{0.55}Se_{0.45}$ [128]. The coexistence of topological phase and BCS-type superconductivity in one iron-chalcogenide material highlights $FeTe_{0.55}Se_{0.45}$ as the connate topological superconductor. In the same system, subsequent STS study captured the zero-energy bound states in the magnetic-vortex cores as the signatures of MBSs [129]. Reminiscent of the tip-induced or intrinsic SC states in Dirac/Weyl semimetals [130], the iron-based superconductors with topological-insulator and/or Dirac-semimetal phases [128,131] similarly combine both superconductivity and topology as the route towards topological superconductivity. In light of the topologically nontrivial phases established in 1-UC FeSe, the discovery of Majorana-like bound states in $FeTe_{0.55}Se_{0.45}$ may stimulate extensive explorations of potential high-temperature topological superconductivity and/or MBSs at 2D limit in 1-UC FeSe [132]. Even without band topology, given the presence of SOC, MBSs for fault-tolerant quantum computation in spin-singlet 1-UC FeSe remain theoretically expectable [133].

## Acknowledgments

The authors acknowledge Zipu Fan for preparing part of the figures, and Ziqiao Wang and Yi Liu for helpful reading of the manuscript. This work is financially supported by National Natural Science Foundation of China (No.11888101), National Key R&D Program of China (No. 2018YFA0305604 and No. 2017YFA0303302), National Natural Science Foundation of China (No. 11774008), Strategic Priority Research Program of Chinese Academy of Sciences (No. XDB28000000) and Beijing Natural Science Foundation (No. Z180010).